\documentclass[conference,a4paper]{IEEEtran}

\usepackage{graphicx, caption, subcaption}% Required for inserting images
\usepackage{algorithm}
\usepackage{array}

\usepackage{algorithmicx}
\usepackage{algpseudocode}

\usepackage{soul}
\usepackage{url}
\usepackage{verbatim}
\usepackage{cite}
\usepackage{listings}
\usepackage[table]{xcolor}
\usepackage[nolist]{acronym}
\usepackage{amsmath}
\usepackage{amssymb}
\usepackage{soul}
\usepackage{rotating}
\usepackage{multirow} 

\usepackage{mathtools, stmaryrd}
\usepackage{xparse} \DeclarePairedDelimiterX{\Iintv}[1]{\llbracket}{\rrbracket}{\iintvargs{#1}}
\NewDocumentCommand{\iintvargs}{>{\SplitArgument{1}{,}}m}
{\iintvargsaux#1} %
\NewDocumentCommand{\iintvargsaux}{mm} {#1\mkern1.5mu..\mkern1.5mu#2}

\usepackage{siunitx}

\usepackage{blindtext}
\usepackage{ctable} % for \specialrule command
\def\BibTeX{{\rm B\kern-.05em{\sc i\kern-.025em b}\kern-.08em
    T\kern-.1667em\lower.7ex\hbox{E}\kern-.125emX}}

\begin{document}

% --------------------------------------------------
% Title
\title{Towards Smart Microfarming \\in an Urban Computing Continuum\\%{\footnotesize \textsuperscript{*}Note: Sub-titles are not captured in Xplore and
}

% --------------------------------------------------
% Autors
\author{\IEEEauthorblockN{%
  Marla Grunewald,
  Mounir Bensalem,  Jasenka Dizdarevi\'c  
   and   Admela Jukan}
\IEEEauthorblockA{%
   Technical University of Braunschweig, Germany\\ \{marla.grunewald, mounir.bensalem, j.dizdarevic,  a.jukan\}@tu-bs.de} %  
%\IEEEauthorblockA{\IEEEauthorrefmark{2}Technical University of Braunschweig, Germany\\
%Email: email@gmail.com}
%\IEEEauthorblockA{\IEEEauthorrefmark{3}Technical University of Braunschweig, Germany\\
%Email: email@gmail.com}
%\IEEEauthorblockA{\IEEEauthorrefmark{4}Technical University of Braunschweig, Germany.}}
}

\maketitle

% --------------------------------------------------
% Abstract
\begin{abstract}
Microfarming and urban computing have evolved as two distinct sustainability pillars of urban living today.  In this paper, we combine these two concepts, while majorly extending them jointly towards novel concepts of smart microfarming  and urban computing continuum. Smart microfarming is proposed with applications of artificial intelligence (AI) in microfarming, while an urban computing continuum is proposed as a major extension of the concept towards an efficient Internet of Things (IoT) -edge-cloud continuum. We propose and build a system architecture for a plant recommendation system that uses machine learning (ML) at the edge to find, from a pool of given plants, the most suitable ones for a given microfarm using monitored soil values obtained from IoT sensor devices. Moreover, we propose to integrate long-distance 
%physical layer 
LongRange (LoRa) communication solution for sending the data from IoT to the edge system, due to its unlicensed nature and potential for open source implementations. Finally, we propose to integrate open source and less constrained application protocol solutions, such as Advanced Message Queuing Protocol (AMQP) and Hypertext Transport Protocol (HTTP) protocols, for storing the data in the cloud. An experimental setup is used to evaluate and analyze the performance and reliability of the data collection procedure and the quality of the recommendation solution. Furthermore, collaborative filtering is used for the completion of an incomplete information about soils and plants.  Finally, various ML algorithms are applied to identify and recommend the optimal plan for a specific microfarm in an urban area. \end{abstract}

% --------------------------------------------------
% Sections
\section{Introduction} \label{sec:intro}
The development of the Internet of Things has revolutionized the many industrial sectors,  and especially in the context of urban living. An urban \emph{pendant} to farming is microfarming, a sustainable farming concept at a smaller scale implemented in urban gardens, typically at less than 5 acres of garden size \cite{jorda2019automated}. Another urban living concept that has hugely benefited from the IoT revolution is urban computing, a pillar of urban sustainability that uses cross-domain data fusion in various context (e.g., air pollution, water quality, traffic) and processes their various modalities (e.g., spatio-temporal and visual) \cite{su15053916}. The voluminous amounts of data that can be collected enable the adaptation of ML and AI tools for different functionalities of use in microfarming. In combination with novel communication and computing technology, including the seamless integration of IoT systems with edge and cloud computing, as well as the availability of free and open source solutions, the sustainable urban development is with no doubt the next innovation frontier. 
The idea to integrate and combine urban microfarming with urban computing, along with AI and ML applications in a computing continuum,  presents an obvious, - but not yet addressed grand challenge. With IoT emerging as a crucial smart agriculture trend, the challenge arises as to how to leverage proprietary system architectures from large scale farming, to seamlessly facilitate an open data exchange between open-source cloud and edge computing systems, and various on-ground devices in urban areas \cite{glaroudis2020survey}.  Especially in the context of the novel and promising computing paradigms, known as computing continuum or IoT/edge/cloud continuum, and their open source implementations, the expected benefits are large in tackling some of the key challenges of urban microfarming \cite{gkonis2023survey}. Put in the context of urban computing, the envisioned \emph{urban computing continuum}, - a novel system architecture that efficiently and seamlessly integrates data, computing and communication in one open system, presents a grand challenge with potential for broader societal impact. We envision a plant recommendation tool to empower urban gardeners to choose the most suitable plants for their gardens and climate conditions.
% our work
%\par 
In this paper, we propose an urban computing continuum architecture and a data processing workflow that enables ML-based plant recommendations in microfarming. The IoT devices include sensors that collect and send data to the edge devices, for data processing, prediction and interface. High computational workloads, such as ML training, are on the other hand executed in the cloud. We study the problem of plant recommendation for microfarming using various ML algorithms, using collected sensor data of soils, as well as its surroundings. We propose that a low-power unlicensed LoRa based communication approach is used for data collection.  The urban computing concept is implemented as a collaborative filtering approach. Assuming a complete knowledge about the performance of plants in different soils, we evaluate the performance of eight different ML techniques and benchmark their performance. The results validate the system performance.  We also show the performance of cosine similarity to predict plant performance in different soils with different data sparsity values. We show that the results carry potential for broader applications also in farming, as experts  traditionally depend on time-consuming and labor-intensive process to obtain conclusions about suitable crops \cite{dahiphale2023smart}. 
% [this should go into conclusion. ]We obtained an accuracy of 93\% with 10\% sparsity and 70\% with 40\% sparsity.  For the plant recommendation results, Gradient Boost (GB) outperforms other approaches with 97.3\% accuracy and 0.06 MSE however it needs 48.6 ms for inference time and 5 second  for training, while Decision Tree  accuracy is 90.2\% and MSE 0.29 with time performance 0.6 ms for inference and 28.7 ms for training.   
%\par 
The rest of the paper is organized as follows. Section \ref{sec:relwork} presents related work. Section \ref{sec:sys} introduces the reference system architecture, including workflow engineering. Section \ref{sec:ml} presents the proposed ML pipeline for the plant recommendation system. Section \ref{sec:results} presents the experimental  and simulation results. Section \ref{sec:conclusion} concludes the paper.

% --------------------------------------------------
\section{Related Work} \label{sec:relwork} %\hl{DONE}

Previous work on microfarming \cite{jorda2019automated} introduces IoT as a pivotal technology, which we further put in the new context of IoT/edge/cloud continuum. In a continuum, IoT devices collect sensor data from fields and send them to an edge device or gateway. For further processing the data is usually forwarded to the cloud. While there are myriad options to use proprietary smart farming software, which carries challenges \cite{ng2021emerging}, IoT/edge/cloud computing concepts today carry potential to offering many open source solutions \cite{jukan2023network}. In terms of communications, LoRa is a long distance protocol, which makes it suitable for smart farming. Paper \cite{9831521} deployed a LoRa based network that uses temperature and humidity sensors to monitor and control irrigation in the field. We adopt a similar approach considering the choice of sensors and information to be obtained from the fields, however the information retrieval does not require using the LoRaWAN infrastructure. Instead, we rely solely physical layer LoRa and offload the computational tasks to the edge based on a open source solution \cite{loramqtt}.
Urban computing is a concept proposed for smart cities to collect and process sensor data to providing residents with intelligent services \cite{su15053916}.  As collected data are prone to incompleteness, collaborative filtering approaches are proposed to predict missing information from collected sensor data  \cite{gao2019collaborative, ortega2020collaborative}.  Sparsity has a high impact of the quality of recommendation system,  thus in \cite{gao2019collaborative} a collaborative filtering recommendation algorithm is based on multi-information source fusion for IoT sensors data.  Similar to previous work,  we will use cosine similarity to our plant recommendation problem in order to predict missing sensor data about the soils. This approach is similar to \cite{10307999}, which explores different  algorithms and expert systems to recommend crops with all data complete and available, whereas our approach is more generic and tolerates data incompleteness. In \cite{9777230}  a single machine learning algorithm, AdaBoost, was used to predict the yield of crops based on  geographic parameters, such as  state, district, area, seasons, rainfall, and temperature. In our paper, we will consider more soil specific features while testing multiple ML algorithms on their efficiency.

% --------------------------------------------------
\section{System Description}\label{sec:sys}

\begin{figure}
    \centering
    \includegraphics[scale=0.85]{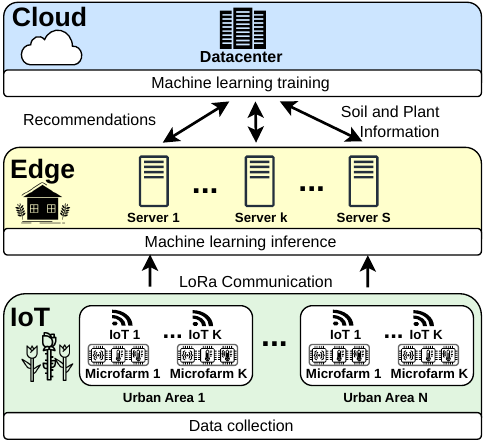}
    \caption{Urban Computing Continuum for Smart Microfarming.}
    \label{fig:architecture}
\end{figure}

\subsection{Reference Architecture}
The IoT/edge/cloud continuum  architecture adopted for a smart monitoring and plant recommendation system in urban microfarming, is illustrated in Figure \ref{fig:architecture}. 
Similar to the concept of urban computing it is designed  to enable an easy deployment of  ML/AI-based applications, thus facilitating the rapid creation and deployment  of models that are efficient to solving problems for urban gardeners. 
The continuum system architecture consists of three types of devices with a range of capabilities that can jointly process, store and communicate data: IoT,  edge, and cloud. Each type of devices has a specific computing capability and constraints as well as different delay, jitter and packet loss along the connections. IoT devices are constrained devices that collect data and send them to the more powerful type of devices, be in the edge or the cloud, such as for storage and processing.    We assume that IoT devices are spread in $N$ urban areas, where each urban area can have one or more microfarms, or gardens, in which the soil characteristics can be monitored by  heterogeneous set of devices, each equipped with a sensor and microcontroller unit. These units are collecting and sending  data on a periodic basis  to an edge server.  This latter can be located in close proximity to the center of an urban area, such as in communal buildings. Each urban area has its own server, resulting in a total of $S$ servers for $N$ urban areas. Communication between the edge and the IoT devices  is assumed using unlicenced long-distance protocols, such as LoRa. The idea behind this choice is on the one hand not to rely on the wireless cellular network, perhaps for monetary reasons, while on the other hand to be able to adopt as many open source solutions as possible \cite{loramqtt}. In the edge computing part of the continuum, machine learning inference is conducted to preprocess data and retrieve insight in the decision making process. While the data collected about soils and plants is stored by the edge servers, a predefined dataset about plants and soils of different microfarms is stored in the cloud and used to train ML models. The trained models are then sent to the edge server to perform the actual recommendation tasks. %

%%%%%%%%%%%%%%%%%%%%%%%%%%%%%%%%%%%%%%%%%%%%%%

% --------------------------------------------------

\subsection{Workflow engineering} 
\begin{figure}
    \centering
    \includegraphics[width=\linewidth]{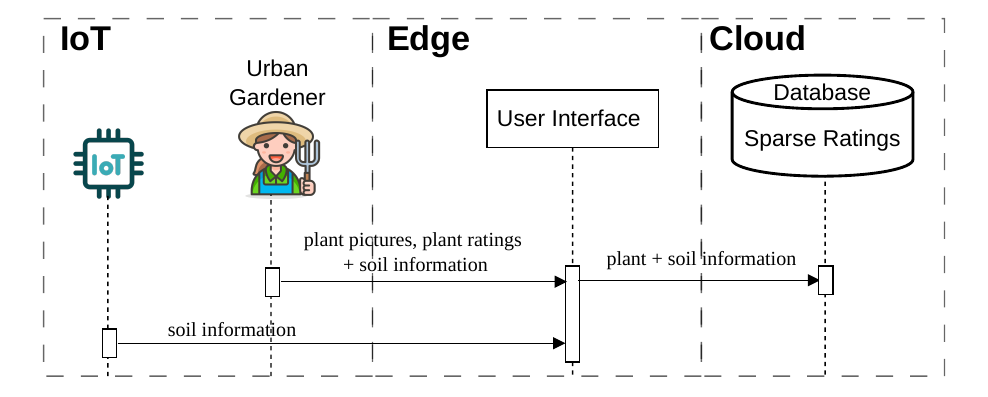}
    \caption{Engineering the microfarming data collection workflow}
    \label{fig:workflow1}
\end{figure}
\begin{figure*}
    \centering
    \includegraphics[width=\linewidth]{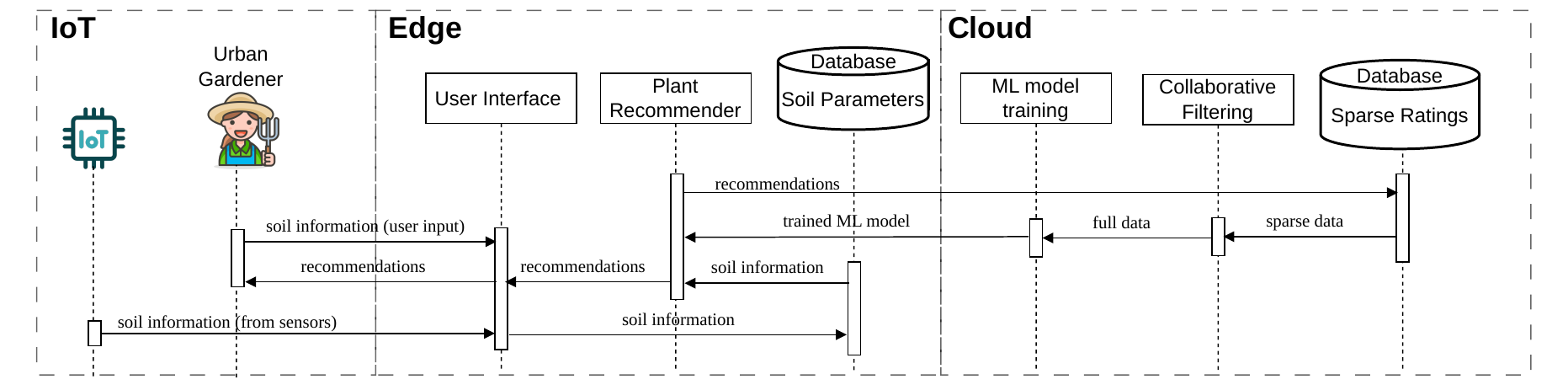}
    \caption{Engineering the microfarming recommendation workflow.}
    \label{fig:workflow}
\end{figure*}
Figure \ref{fig:workflow1}  illustrates the engineered microfarming \emph{data collection workflow} for the proposed system architecture. In the front-end application, the urban gardener initiates the entire process as a user (system activation). This is accomplished through the user interface, which can be built with various degrees of sophistication (e.g., video, text, social media links). The activation of the system initiates the collection of data from sensors located in the garden's soil, in the IoT context of the continuum. The sensors measure the concentration of the essential macro nutrients \textit{nitrogen (N)}, \textit{phosphorus (P)}, and \textit{potassium (K)} as well as the \textit{temperature} and \textit{pH }value;  additional parameters can be collected from diverse sources (e.g., air quality, water quality, environmental data). As each sensor is interfacing a microcontroller, this information is then written internally to a file and stored for further processing in the edge computing context. These files with sensor measured parameters are then forwarded to the edge server, via a communication protocol that is suitable for the required distance between the field and the edge server, which includes a database that stores the soil parameters of the microfarm's soil. The urban gardener is assumed to manually provide ratings to be stored in the cloud to create a dataset of plants and their ratings under \emph{Sparse Ratings}. On the other hand, the gardener could also send plants pictures for automated rating through ML models, which is out of scope here.  \par Figure \ref{fig:workflow} illustrates the engineered microfarming \emph{recommendation workflow}, where  the urban gardener  initiates the recommendation process as a user by  providing the soil information either manually or through sensors to the user interface. In order to recommend plants, we first process the ratings dataset in the cloud.  The sparse nature of these data is due to the fact that not all plants are grown by one gardener in the same garden, or some gardeners do not enter all the data, thus resulting in a significant number of missing values of ratings within the database. In order to complete the existing database, we apply collaborative filtering algorithms to address the gaps in the ratings, thereby creating a complete ratings file for the urban microfarm ecosystems. This file is then utilized for training of the ML models, also in the cloud context, with the input for the ML models consisting of the soil parameters, while the output determines the ratings. The model gets retrained periodically, with the time period $T$ being chosen by the administrator of the entire system, based on its scale. The trained model is then moved to the so-called \emph{Plant Recommender} of the edge context where it is run using the current soil parameters from the database collected from the farms and as a result provides a recommendation to the urban gardener. Additionally the recommendation is added to the file of the full ratings to enlarge the dataset. 
%%%%%%%%%%%%%%%%%%%%%%%%%%%%%%%%%%%%%

\section{ML pipeline for plant recommendation}\label{sec:ml}
\subsection{Data collection}
The proposed system employs a heterogeneous set of IoT devices, deployed by using low power resource constrained hardware devices (e.g., microcontrollers) each interfacing a different sensor. The microcontrollers that each of the sensors is interfaced with are typically resources constrained devices and are expected to be located in remote areas which are connected over LoRa based communication, operating over a license-free spectrum and is low power, which makes it another sustainable solution.  To achieve this, each microcontroller is equipped with a LoRa transmitter module, which allows it to send sensor data as LoRa packets. For these messages to reach the edge device (e.g., Raspberry Pi), it also has to be connected with a LoRa receiver module.  Here, several  communication parameters should be taken into account and configured on both receiver and transmitter side, such as  the spreading factor, coding rate, transmission frequency and  bandwidth. %In Germany,  the  transmission frequency can be set to 870 MHz, as this frequency does not impose duty cycle restrictions, provided that the transmission power is below 500 mW  \cite{FreqPlan} \hl{not sure why just this parameter is mentioned, should be all here or none}.  
 Once the LoRa packets with sensor data reach the edge device, they are extracted for data cleaning and preprocessing. In addition, the extracted data is also encapsulated into messages of less constrained communication protocols, such as HTTP or AMQP, which are then used to store the data from the edge device to the cloud. The individual soil values sensors gather directly from the soil are as follows: \textit{N}, \textit{P},  \textit{K},  \textit{temperature} and \textit{pH } value, while further parameters can be measured from the environment. 
 
\subsection{Pre-processing}
The pre-processing of the data is transforms the sparse matrices to be complete, to be made suitable for training ML models. This is achieved by calculating the missing ratings from the sparse matrix using cosine similarity, thus generating a complete ratings matrix  for all existing plant and soil combinations. In the context of recommendation systems, the cosine similarity metric is calculated using the ratings of two items, designated as x and y, which represent two rows in the sparse matrix \cite{cosinesimbooks}. Our input data represents the ratings on each soil, and the cosine similarity is given by:
\begin{equation}
\cos(x,y) = \frac{x \cdot y}{\|x\| \|y\|}
\end{equation}

The sparse matrix $S \in \mathbb{N}^{m \times n}$ is composed of natural numbers $ \in \Iintv{1,5}$, with 1 being the lowest rating and 5 the highest, as well as missing ratings in between. The parameters $m$ and $n$ describe the amount of soil samples and the number of plants that can be rated, respectively. Using cosine similarity on the dataset to compare all $m$ rows with each other, and the full matrix $F \in \mathbb{N}^{m \times n}$ is build.
  Any missing values are filled in the sparse matrix by identifying similar soils based on the similarity matrix and calculated the weighted average of their ratings based on the similarity. These ratings are then added to the sparse matrix to create a complete data set of ratings.
\subsection{Recommendation approaches} %\hl{Done}
After collection of soil data, and after pre-processing data into ratings between 1 and 5, we apply cosine similarity to obtain full information about the rating of each plant in each existing soil.  Afterwards, the actual recommendation can be accomplished through the application of machine learning models using the full ratings 2-dimensional matrix as input. We  adopt several widely available ML algorithms to be used for plant selection, which will be benchmarked  and compared in terms of accuracy, error rate, training time and inference time. The adopted ML algorithms are: K-Nearest Neighbors (KNN), Neural Networks (NN),   Linear Regression, Decision Tree, Random Forest, Support Vector Machine (SVM), Gradient Boost (GB), and Extreme Gradient Boost (XGB). Those algorithms are applied to recommend the $N$ best plants suitable for a given microfarm soil.

%%%%%%%%%%%%%%%%%%%%%%%%

\section{Experimental and Simulation Results}\label{sec:results} 

This section evaluates various aspects of the proposed system, including the efficiency of low-power communication solution (LoRa) for sending sensor data from the IoT devices to the edge, the use of collaborative filtering for the completion of incomplete sensor information and crop recommendation system based on various ML algorithms. The experiment and simulation parameters are listed in Table \ref{tab:table2}. 

\begin{table}[h!]
  \begin{center}
    \caption{Simulation Parameters}
    \label{tab:table2}
    \begin{tabular}{|c|c|c|}
     \specialrule{.15em}{0em}{0em} % \hline 
      %\textbf{Value 1} & \textbf{Value 2} & \textbf{Value 3}\\
      \textbf{Experiment/Simulation  } &    \textbf{Parameters} & \textbf{Values} \\

      \specialrule{.15em}{0em}{0em} % \hline
      LoRa & Distance & 35 m \\
      (Experiment) & Frequency  & 870 Mhz \\
        & Bandwidth  & 125 kHz \\
        & Coding Rate  & 4/5 \\
        & Spreading Factor  & 7 \\
      \hline
       Collaborative  &  library  & scikit-learn  \\
        filtering &    & $\rightarrow$ cosine\_similarity \\
        (Simulation)   & Number of plants  & 15 \\
         & Number of soils  & 10626  \\  
         \hline
       ML  recommendation & Model library  & scikit-learn \\
        (Simulation) & Number of soils  & 10626 \\
         & Number of plants  & 15 \\
          & test size data  & 20 \% \\
      \specialrule{.15em}{0em}{0em} % \hline\hline
    \end{tabular}
  \end{center}
\end{table} 

%--------------------------------------------------
\subsection{LoRa performance}

\begin{table}[!ht]
	\label{tableExp}

	\center
	\renewcommand{\arraystretch}{1} 
	\caption{Mean RSSI and SNR values for tested scenarios}
	\setlength{\tabcolsep}{3pt}
	\begin{tabular}{cccccc}
		
		\toprule
    \textbf{Sc.}   &  \textbf{CAD}   &  \textbf{Packets}    & \textbf{Payload}       &    \textbf{Mean RSSI} & \textbf{Mean SNR} \\
    &   &  \textbf{received}    & \textbf  &   &  \\
	\midrule
  1  & -   & $100$ \%   & $3$ B   &    -$75$ dBm & $11$ dB \\
  2  & Yes   & $100$ \%, $100$ \%    & $3$ B    & -$96$ dBm, -$90$ dBm & $5$ dB, $9$ dB \\
  3  &  No   & $54$ \%, $88$ \%   & $3$ B   &    -$96$ dBm, -$72$ dBm & $5$ dB, $11$ dB \\
	\midrule
	  
  1  &-   & $100$ \% & $50$ B   &    - $88$ dBm &  $10$ dB \\

  2  & Yes   & $100$ \%, $100$ \%    & $50$ B    & -$61$ dBm, -$78$ dBm & $11$ dB, $11$ dB \\
  3  &  No   &$10$ \%, $46$ \% & $50$ B   &    -$77$dBm, -$61$ dBm & $12$ dB , $10$ dB \\
	\midrule
  1  &	 -   & $100$ \%  & $250$ B   &    - $76$ dBm &  $11$ dB\\
	
  2  &  Yes   & $100$ \%, $100$ \%    & $250$ B    & -$68$ dBm, -$79$ dBm & $12$ dB, $11$ dB\\
  3  &	 No  & $0$ \%, $17$ \%   & $250$ B   &    /, -$65$ dBm & /, $10$ dB \\
	
	\bottomrule
	\end{tabular}
	\label{tab:tab1}
\end{table}
%\end{sidewaystable*}

Table \ref{tab:tab1} presents the data obtained from sending data from an IoT sensor device to the edge device, using LoRa transmission, with the experiments conducted in the laboratory. %To optimize the LoRa communication parameters to a potential use with low power energy sources, its transmission parameters, including bandwidth, spreading factor and coding rate were set to 125 kHz, 7 and 4/5, respectively. The transmission frequency used was set to 870 MHz, as this frequency does not impose duty cycle restrictions in Germany, provided that the transmission power is below 500 mW \cite{FreqPlan}. 
The experiments were conducted by transmitting 100 LoRa packets with a 5-second interval between each transmission. The measured values include the percentage of received packets, the Received Signal Strength Indicator (RSSI) as well as the Signal to Noise Ratio (SNR). 
%Three scenarios were tested to show the impact of interference on data transmission from multiple IoT devices. 
In Scenario 1, a single IoT device transmitted data via LoRa, serving as the control case. In Scenario 2, two devices transmitted simultaneously, with a Collision Avoidance Detection (CAD) mechanism activated. This mechanism involved waiting a randomly generated time interval of less than two seconds to check if the channel was available. If the channel was busy, the system would wait before checking again and then transmit when the channel was free. In Scenario 3, two devices also transmitted simultaneously, but the CAD mechanism was deactivated to demonstrate the effects of interference on data transmission from multiple IoT devices. For the experiments, the payload lengths were set to 3 , 50 and 250 bytes, respectively, to illustrate a low, medium and high payload value. 

%Three scenarios where observed, in order to evaluate the influence of interference when there is more than IoT device sending the data, one (scenario 1) with a LoRa transmission of a single IoT device, while the other two with two devices transmitting simultaneously (scenario 2 and 3), resulting in packet collisions measured in the edge device. To prevent collisions, a Collision Avoidance Detection (CAD) mechanism is implemented and activated for scenario 2. This mechanism waits for a randomly generated time interval that is shorter then two seconds before checking if the channel is currently occupied. If the channel is still occupied, it waits again before rechecking and eventually transmits the data if the channel is free. For scenario 3, this mechanism is deactivated to demonstrate the impact of collisions. For the experiments, the payload lengths were set to 3 and 250 bytes, respectively, to illustrate a low and high payload value. 

 The results of the first scenario demonstrate that, in absence of an interfering second LoRa transmission, irrespective of the payload length, all transmitted packets were successfully received. Furthermore, the RSSI and SNR values exhibit comparable ranges in this scenario. The results of the second scenario demonstrate that with an activated CAD, 100\% of the packets are received for all three payload lengths from both sending devices. The RSSI and SNR values deviate from those observed with a single transmitting device, exhibiting a range of variation both above and below the mean. The third scenario demonstrated the influence of the CAD. In this case, it is shown that in the three byte scenario, 54\% of the packets were received from one of the devices, while 88\% were received from the other, showing a significant loss. 

%The RSSI and SNR values are  greater for the sending device with a higher number of received packets \hl{this has to be explained why}.  
The RSSI and SNR values are observed to be higher for transmitting devices with a greater number of received packets. This phenomenon can be described in terms of the capture effect, whereby a receiver is still able to demodulate the packet with the higher RSSI in the event of a packet collision \cite{capture}.
With the larger payload sizes of the sensor data, the amount of packets received in the edge device decreases significantly in the third scenario, without the activated CAD (0\%  and 17\% of the packets were received for 250B payload, compared to  54\% and 88\% for 50B payload).  This phenomenon can be attributed to the fact that the transmission time is also lengthening due to the necessity of transmitting a greater number of symbols, thereby increasing the probability of two signals colliding and one being extinguished by the other. 
The results of the experiment indicate that, in the context of a single transmitting device, the payload length has a negligible impact on the performance of LoRa transmissions. However, once more than one active LoRa device is attempting to send to a common receiver, packet collision might appear causing messages to be lost. To prevent this, a CAD mechanism should be implemented to ensure the reliability of the system. 

% -------------------------------------------
\subsection{Collaborative filtering performance}% \hl{reviewing here ..}
\begin{figure*}
 \centering %\vspace{-0.5 cm}
 \begin{subfigure}[t]{0.24\linewidth}
 \includegraphics[scale=0.20]{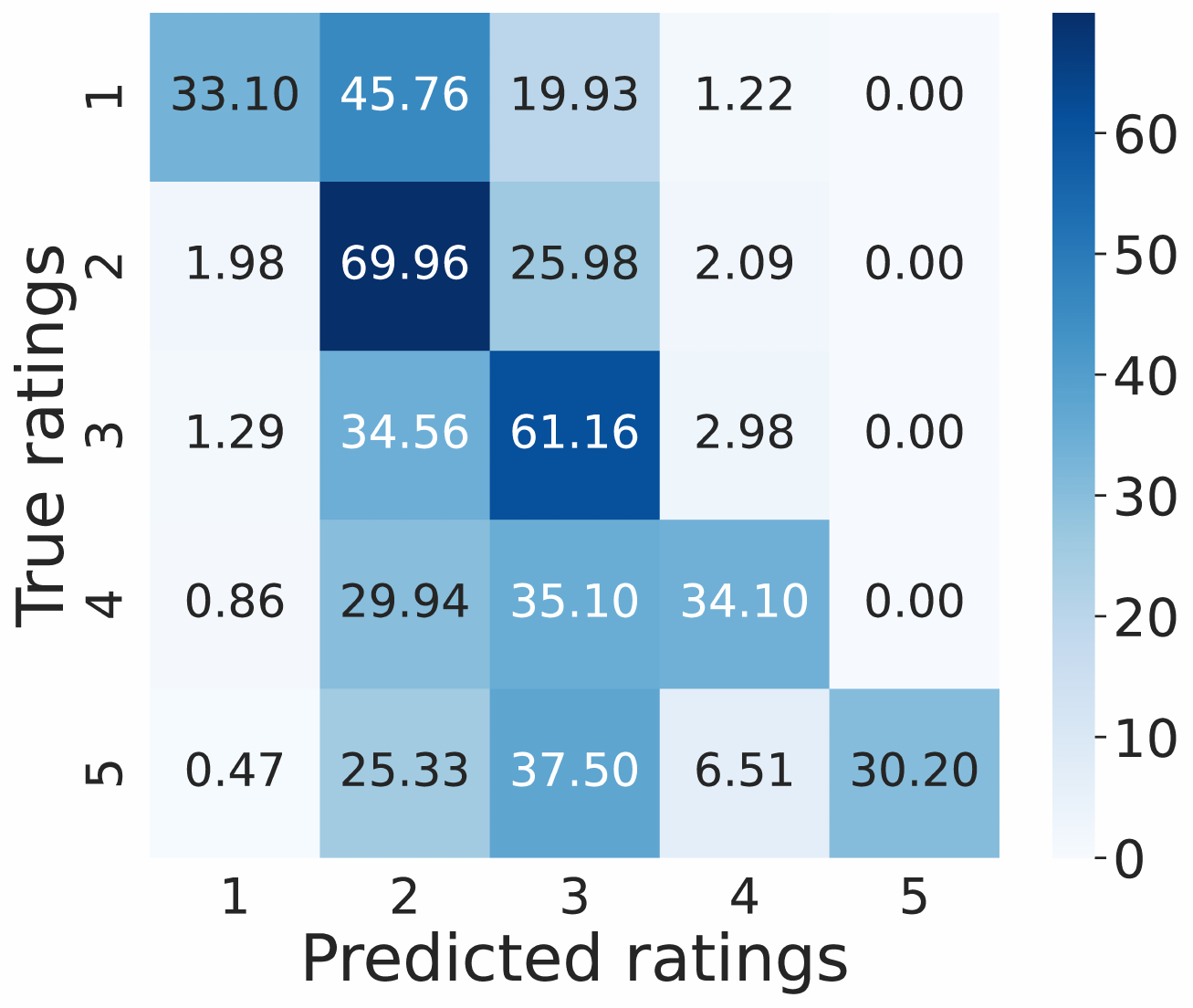}
  \caption{Sparsity=70\%  }\label{subfig:mat1}
   \end{subfigure}
   \begin{subfigure}[t]{0.24\linewidth}
 \includegraphics[scale=0.20]{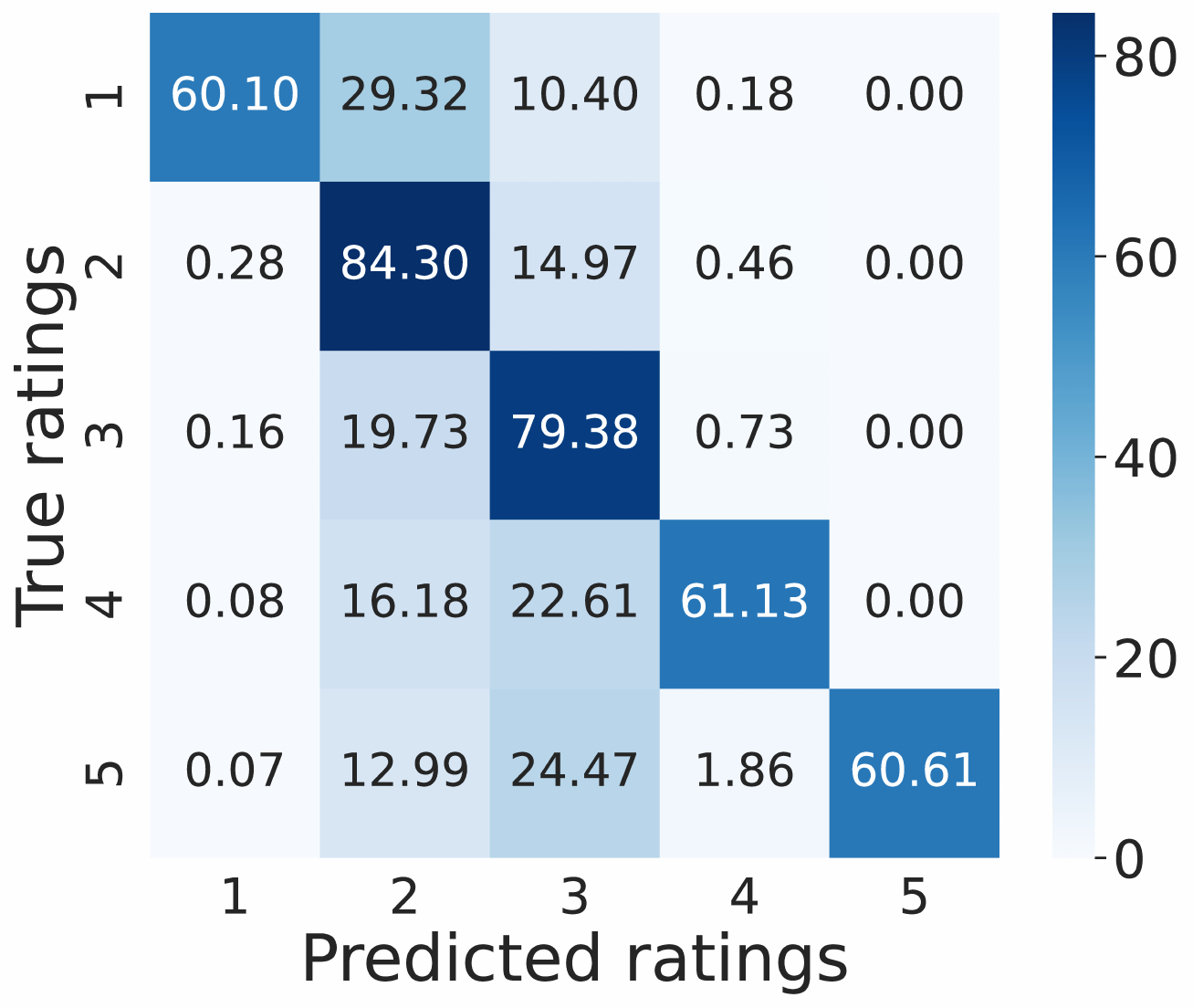}
 \caption{Sparsity=40\% }\label{subfig:mat2}
   \end{subfigure}
 \begin{subfigure}[t]{0.24\linewidth}
 \includegraphics[scale=0.20]{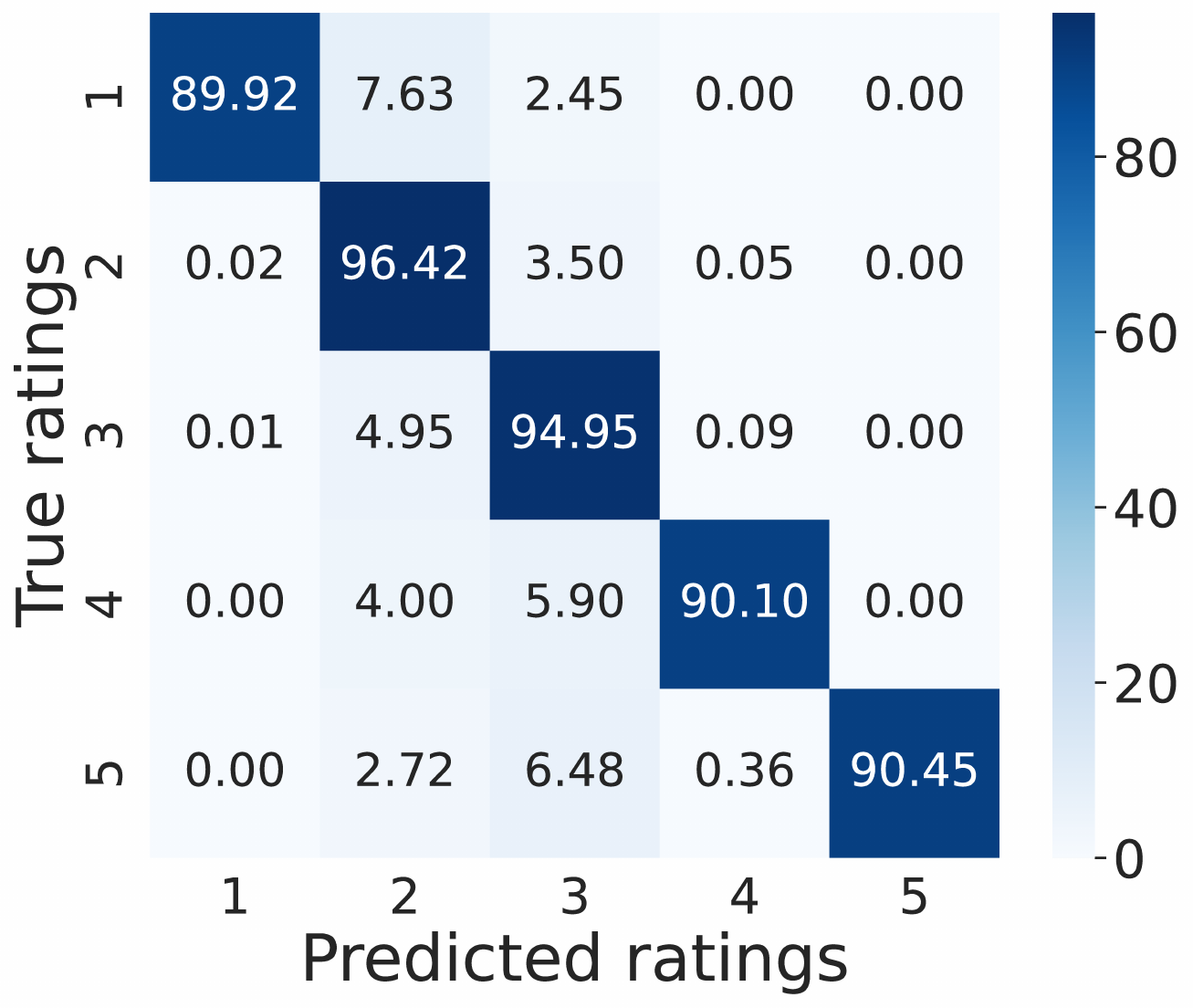}
 \caption{Sparsity=10\%  }\label{subfig:mat3}
   \end{subfigure}
    \begin{subfigure}[t]{0.24\linewidth}
 \includegraphics[scale=0.20]{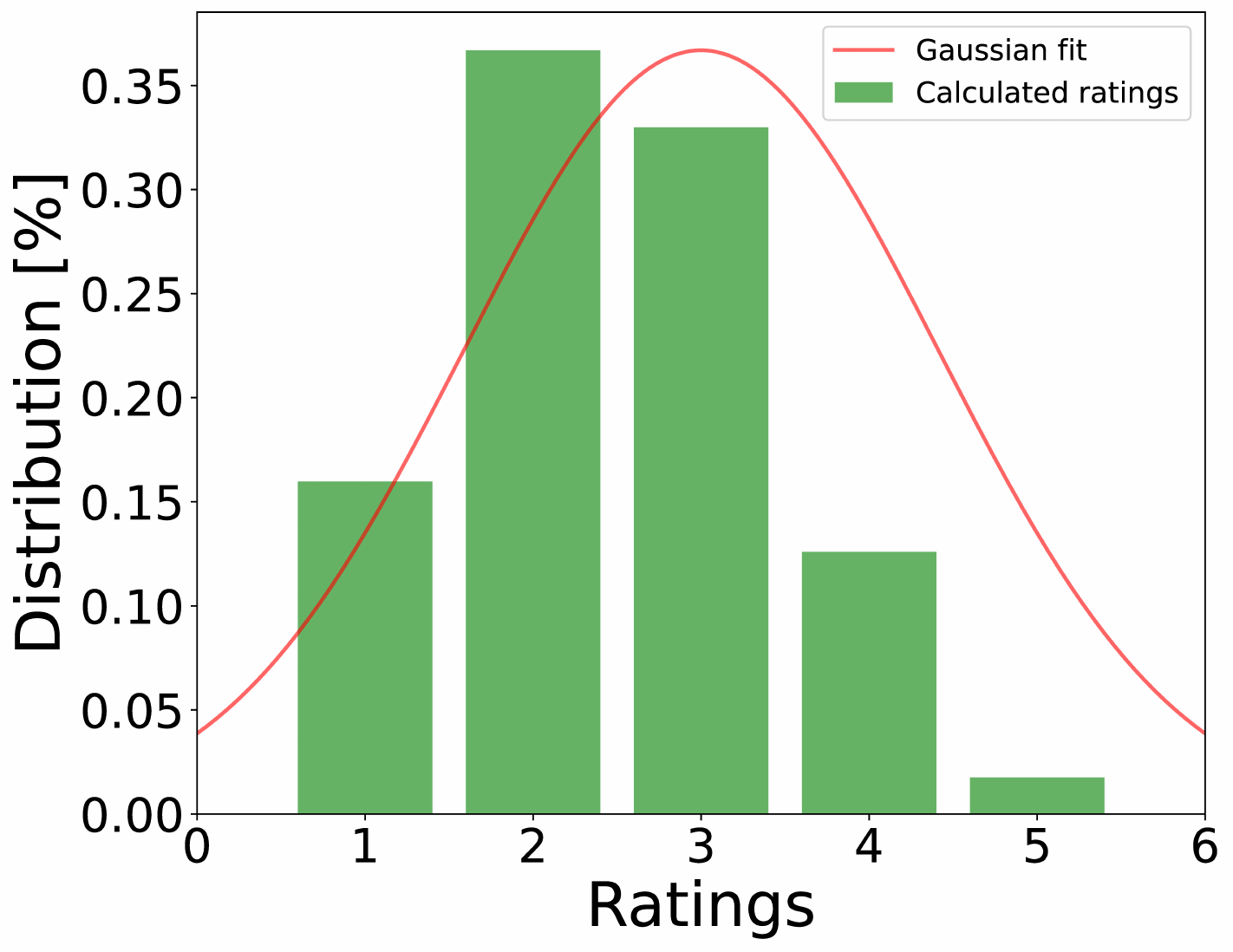}
 \caption{Rating distribution  }\label{subfig:ratings}
   \end{subfigure}
 \caption{Confusion matrix with different data sparsity and rating distribution.}
\label{fig:confmat}
\vspace{-0.2 cm}
\end{figure*}

The performance of the collaborative filtering method, i.e., cosine similarity,  is shown in Figure \ref{fig:confmat}. The input data is represented as a 2D matrix of ratings of plants in different soils. 
To compute the performance of the cosine similarity the two datasets get compared using a confusion matrix to show how accurate the prediction of the rating has been done. In order to demonstrate the impact of sparsity on the performance, this experiment has been conducted with varying levels of sparsity. Sparsity is defined as the ratio of abstinent ratings to the total number of ratings. Three distinct sparsity levels were tested: 70\% , 40 \% and 10 \%. The results are presented in Figures \ref{subfig:mat1} to \ref{subfig:mat3}. We can observe that as sparsity decreases, the performance improves, resulting in more precise predictions. A sparsity of 70 \% based on a dataset comprising 10626 soils and 15 plants in insufficient for the reliable prediction of ratings. The performance improves with an increase in sparsity to 40 \%. In this experiment, the performance for the outlier ratings (1,4,5) was satisfactory. However, confusion was evident for the ratings 2 and 3. This could be explained by referring  to Figure \ref{subfig:ratings}. The distribution of total ratings is not uniform. The majority of ratings are in the range of ratings 2 and 3, which presents a challenge for the algorithm in distinguishing between them. This issue is mitigated when the sparsity is reduced to 10 \%. In this case, the data set is sufficient to reliably calculate the correct ratings, although the performance for ratings 2 and 3 remains slightly poorer than for the other three ratings.

%\begin{table}[h!]
%    \centering
%    \begin{tabular}{ cc|c|c|c|c|c|c } 
%  & \multicolumn{6}{c}{Prediction} \\ 
%\multirow{7}{*}{\rotatebox{90}{True Ratings}} &  & 1 & 2& 3 & 4& 5\\ \cline{2-7}
%    & 1 &  \cellcolor{gray!97}{\textbf{97.05}}  & \cellcolor{gray!3}{\textbf{2.32}}  & \cellcolor{gray!0}{0.63} & \cellcolor{gray!0}{0}& \cellcolor{gray!0}{0}\\ \cline{2-7}
%    & 2 &   \cellcolor{gray!5}{\textbf{4.69}}    &\cellcolor{gray!95}{\textbf{95.15}}  & \cellcolor{gray!0}{0.16} & \cellcolor{gray!0}{0}& \cellcolor{gray!0}{0}\\ \cline{2-7}
%    & 3 &  \cellcolor{gray!0}{0.6}   & \cellcolor{gray!6}{\textbf{5.52}}   &  \cellcolor{gray!93}{\textbf{93.89}}  &\cellcolor{gray!0}{0}& \cellcolor{gray!0}{0}\\ \cline{2-7}
%    & 4 &  \cellcolor{gray!0}{0} & \cellcolor{gray!1}{\textbf{0.89}} &  \cellcolor{gray!7}{\textbf{6.45}}  & \cellcolor{gray!92}{\textbf{92.66}}  &  \cellcolor{gray!0}{0}\\ \cline{2-7}
%    & 5  & \cellcolor{gray!0}{0} & \cellcolor{gray!0}{0} &  \cellcolor{gray!1}{1.05}&  \cellcolor{gray!6}{\textbf{6.09}}  &  \cellcolor{gray!92}{\textbf{92.86}}\\ \vspace{0.1pt}
%\end{tabular}
%    \caption{Caption}
%    \label{tab:my_label}
%\end{table}

\subsection{Plant recommendation} 
\begin{figure}
 \centering %\vspace{-0.5 cm}
   \includegraphics[scale=0.43]{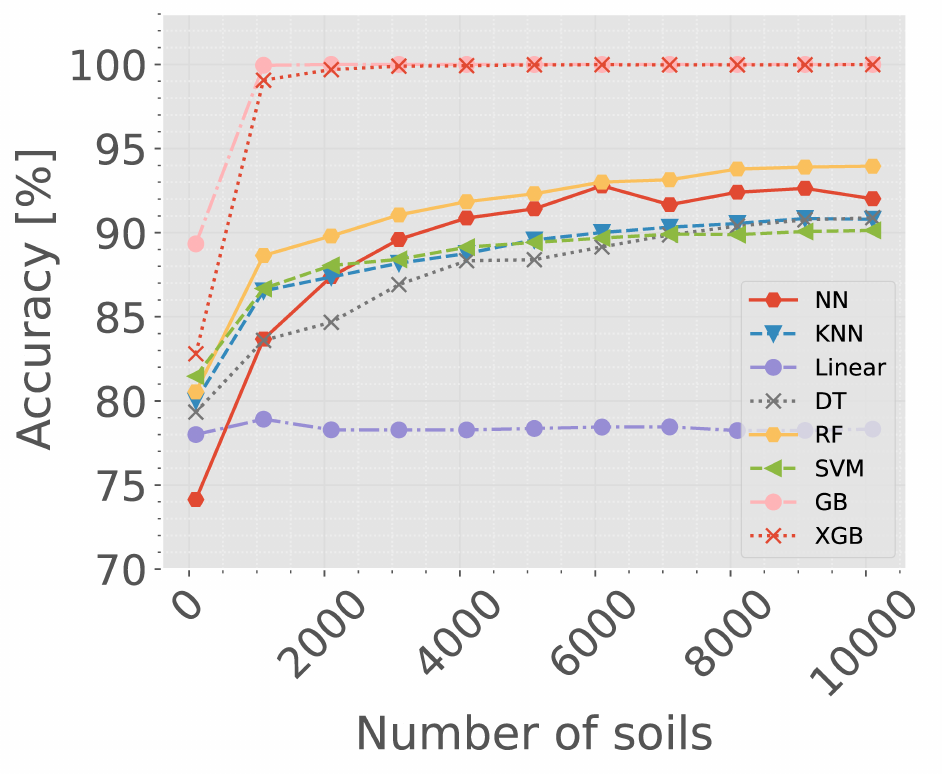}
 \caption{Prediction performance vs number of soils}
\label{fig:accvssoils}
\vspace{-0.3 cm}
\end{figure}

The results of the plant recommendation are presented in Table \ref{tab:algo} and Figure \ref{fig:accvssoils} with evaluated metrics of accuracy, mean square error (MSE), the model training and inference time. For each algorithm, the model was tested with data sets of varying sizes, ranging from 100 to 10100 soils and their corresponding ratings, in increments of 1000. Figure \ref{fig:accvssoils} illustrates the change in accuracy of the eight algorithms utilized as the number of soils in the data set increases.  Gradient Boost (GB) and Extreme Gradient Boost (XGB) approach 100 \% quickly. In contrast, the linear algorithm exhibits a relatively constant performance, maintaining an accuracy level of approximately 78\% over all tested sample sizes. The remaining five algorithms demonstrate a similar trend, with an increasing accuracy as the soil number increases. Table \ref{tab:algo} presents the precise values for the four tracked metrics at the highest tested soil count  $\sim$ 10$^4$. The highlighted numbers indicate the highest values for the respective metric. Overall, with the exception of the linear algorithm, the accuracy reaches 90 \% or higher, with the highest value of 99.99 \% measured for GB. The best value for the MSE was observed for the XGB, with a value of 0.0003. 
The sole exception is the linear algorithm, which exhibits an exceedingly poor MSE of 0.65. The third metric describes the time required for model training in the cloud. These results demonstrate that the high accuracy values are accompanied by extended training periods. The only algorithm that requires a longer inference time than training time is KNN. This is due to the fact that no actual learning occurs during the training phase of KNN, this only occurs during the inference phase, which explains the discrepancy in time requirements. 
Based on these results, an algorithm can be selected that is appropriate for this use case. In the absence of time constraints, the decision can be made exclusively on the basis of accuracy and MSE. Consequently, the two algorithms GB and XGB are the preferred options.

%\begin{table*}  % Old results
%\centering
%\begin{tabular}{SSSSSSSSSS} \toprule
%    {\textbf{Metrics}} & {\textbf{NN}} & {\textbf{KNN}} & {\textbf{Linear}} & {\textbf{DT}} & {\textbf{Random Forest}} & {\textbf{SVM}}& {\textbf{Adaboost}} & {\textbf{Gradient Boost}}  & {\textbf{Extreme Gradient Boost}} \\ \midrule
%  {\textbf{F1 score [\%]}} & {91.1} & {94.1} &{89.1} & {96.9} &{91.5} & {89.7} &{94.1} & {\textbf{99.2}} & {87.2} \\ \midrule
% {\textbf{Accuracy [\%]}}& {86.1} & {90.4} &{81.8} & {94.8} &{86.7} & {84.1} &{90.4} & {\textbf{98.8}} & {80.9} \\\midrule
% {\textbf{MSE} } & {0.11} & {0.18} & {0.66} & {0.20} & {0.10} & {0.18} & {0.17} & {0.002} & {\textbf{0.0002}}\\ \midrule
%    {\textbf{Training time [ms]}}  & {6526.6} & {3.9}  &{\textbf{1.3}} & {33.9} &{2145.3} & {30771.6} &{6763.9} & {5941.7} & {367.9} \\ \midrule
%    {\textbf{Inference time [ms]}}  & {0.9} & {13.7}  &{\textbf{0.1}} & {0.6} &{46.6} & {6378.1} &{162.9} & {51.4} & {16.5}  \\ \bottomrule
    
%\end{tabular}
%\caption{Algorithms performance.\hl{To be filled}}
%\label{tab:algo}
%\end{table*}

%-------------------------------------------------------------------------------

\begin{table*}
\centering
\begin{tabular}{SSSSSSSSS} \toprule
    {\textbf{Metrics}} & {\textbf{NN}} & {\textbf{KNN}} & {\textbf{Linear}} & {\textbf{Decision Tree}} & {\textbf{Random Forest}} & {\textbf{SVM}} & {\textbf{Gradient Boost}}  & {\textbf{Extreme Gradient Boost}} \\ \midrule
%  {\textbf{F1 score [\%]}} & {92.5} & {90.3} &{78.1} & {90.2} &{93.3} & {89.6} &{83.5} & {\textbf{97.3}} & {96.8} \\ \midrule
 {\textbf{Accuracy [\%]}}& {92.0} & {90.8} &{78.3} & {90.9} &{94.0} & {90.1} & {\textbf{99.99}} & {99.9} \\\midrule
 {\textbf{MSE} } & {0.15} & {0.19} & {0.65} & {0.26} & {0.12} & {0.19} & {0.002} & {\textbf{0.0003}}\\ \midrule
    {\textbf{Training time [ms]}}  & {6960.1} & {3.0}  &{\textbf{1.1}} & {22.2} &{1455.0} & {15646.3} & {4288.4} & {314.4} \\ \midrule
    {\textbf{Inference time [ms]}}  & {0.9} & {9.6}  &{\textbf{0.1}} & {0.4} &{30.2} & {2997.4} & {34.6} & {14.1}  \\ \bottomrule
    
\end{tabular}
\caption{Algorithms performance.}
\label{tab:algo}
\end{table*}

%--------------------------------------------------
\section{Conclusion}\label{sec:conclusion} 
In this study, we presented a comprehensive IoT-edge-cloud architecture and an integrated data workflow tailored for ML-based applications in smart farming and plant recommendation. The architecture comprises an IoT layer with sensors for data collection, an edge layer with dedicated servers for data processing, prediction, and user interaction, and a cloud layer for handling high computational workloads such as ML training. We addressed the plant recommendation problem by employing various ML algorithms on sensor data, including soil composition and climate data.
Our proposed communication approach leverages LoRa for efficient data collection, and we introduced a collaborative filtering method using cosine similarity to handle missing data.  Our findings indicate that the payload length of LoRa messages does not impact connection reliability, though a collision avoidance mechanism is essential when multiple LoRa devices are employed to prevent packet loss.  The evaluation demonstrated that cosine similarity effectively predicts plant performance in varying soil conditions, achieving an accuracy of 93\% with 10\% data sparsity and 70\% with 40\% data sparsity.   Assuming complete knowledge of plant performance in diverse soil conditions, we evaluated eight ML techniques for crop recommendation.
Among the ML techniques, Gradient Boost (GB) achieved superior performance with 99.99\% accuracy and 0.002 MSE, requiring 34.6 ms for inference and 4.3 seconds for training.
%In contrast, the Decision Tree attained 90.9\% accuracy and 0.26 MSE, with significantly faster times of 0.4 ms for inference and 28.7 ms for training.
% Limitation
One key limitation is the reliance on a single communication protocol, LoRa, which, despite its efficiency, may face scalability and interference challenges in densely populated sensor environments. 
% Future Work
Future research will focus on optimizing ML algorithms for deployment on edge devices, through techniques such as model compression and hardware acceleration, will be a priority to achieve faster inference times without compromising accuracy. 

%\blindtext[2]

% --------------------------------------------------

\section*{Acknowledgment}
This work was partially supported by the  EU project MANOLO  under GA no. 101135782.

% --------------------------------------------------
% References
\bibliographystyle{IEEEtran}
\bibliography{references_2}

\end{document}